\documentclass[aps,twocolumn,pra,superscriptaddress,tightenlines]{revtex4}
\usepackage{amsmath}
\usepackage{amssymb}      
\usepackage{amsfonts}
\usepackage{graphicx}
\usepackage{color}
\usepackage{xcolor}
\usepackage[colorlinks=true,linkcolor=blue,citecolor=blue,urlcolor=blue]{hyperref}
\usepackage{booktabs}
\begin{document}
	
	\title{Floquet dark-mode manipulation in a three-mode optomechanical system}
	\author{Mian Mao}
	\affiliation{Key Laboratory of Low-Dimensional Quantum Structures and Quantum Control of Ministry of Education, Key Laboratory for Matter Microstructure and Function of Hunan Province, Department of Physics and Synergetic Innovation Center for Quantum Effects and Applications, Hunan Normal University, Changsha 410081, China}
	\affiliation{Hunan Research Center of the Basic Discipline for Quantum Effects and Quantum Technologies, Hunan Normal University, Changsha 410081, China}
	\author{Jin-Feng Huang}
	\email{Contact author: jfhuang@hunnu.edu.cn}
	\affiliation{Key Laboratory of Low-Dimensional Quantum Structures and Quantum Control of Ministry of Education, Key Laboratory for Matter Microstructure and Function of Hunan Province, Department of Physics and Synergetic Innovation Center for Quantum Effects and Applications, Hunan Normal University, Changsha 410081, China}
	\affiliation{Hunan Research Center of the Basic Discipline for Quantum Effects and Quantum Technologies, Hunan Normal University, Changsha 410081, China}
	\affiliation{Institute of Interdisciplinary Studies, Hunan Normal University, Changsha 410081, China}
	\author{Jie-Qiao Liao}
	\email{Contact author: jqliao@hunnu.edu.cn}
	\affiliation{Key Laboratory of Low-Dimensional Quantum Structures and Quantum Control of Ministry of Education, Key Laboratory for Matter Microstructure and Function of Hunan Province, Department of Physics and Synergetic Innovation Center for Quantum Effects and Applications, Hunan Normal University, Changsha 410081, China}
	\affiliation{Hunan Research Center of the Basic Discipline for Quantum Effects and Quantum Technologies, Hunan Normal University, Changsha 410081, China}
	\affiliation{Institute of Interdisciplinary Studies, Hunan Normal University, Changsha 410081, China}
	\date{\today }

	\begin{abstract}
		
			We propose a Floquet sideband method to manipulate the dark-mode effect in a three-mode optomechanical system consisting of a single cavity mode coupled to two mechanical modes. We introduce a periodic frequency modulation for one mechanical mode to generate the Floquet sidebands, one of which is selected to match the resonance frequency of the other mechanical mode, allowing complete destructive interference to produce a mechanical dark mode decoupled from the cavity mode. The semiclassical operating point is determined self-consistently within the framework for linearization of optomechanical interaction, and the linear stability is examined under the rotating-wave approximation. We evaluate the validity of the approximate Hamiltonian by examining the fidelity between the states governed by the approximate and exact Hamiltonians. We further demonstrate dark-mode control by inspecting the bright/dark-mode populations of the system, when a frequency detuning is introduced suddenly. These results may provide a basis for further study of quantum-state storage using the Floquet mechanical dark modes.
			
	\end{abstract}
	
	\maketitle
	
	\section{Introduction}
		
    Cavity optomechanics \cite{Kippenberg2008,Aspelmeyer2014} addresses the radiation-pressure interactions between the electromagnetic fields and the mechanical motion of the moving boundary \cite{Law1995}. Over the past two decades, great advances have been achieved in understanding and manipulating optomechanical interactions. In particular, advances in mechanical cooling \cite{Gigan2006,Arcizet2006,Kleckner2006,Rocheleau2010,WilsonRae2007,Marquardt2007,Teufel2011,Chan2011,Lai2018,Sommer2019,Naseem2021}, strong coupling \cite{Dobrindt2008,Groblacher2009,Teufel2011b,Verhagen2012}, optomechanically induced transparency \cite{Weis2010,SafaviNaeini2011,Agarwal2010}, and coherent state transfer \cite{Palomaki2013} have established key capabilities for optomechanical quantum manipulation, including quantum sensing \cite{Metcalfe2014,Schreppler2014,Mason2019}, frequency conversion \cite{SafaviPainter2011,Hill2012,Bochmann2013,Andrews2014,OckeloenKorppi2016}, quantum communication \cite{Stannigel2010}, and quantum information processing \cite{Stannigel2012}. Extending optomechanical manipulation to interactions and correlations between mechanical oscillators motivates the study of multimode optomechanical systems \cite{Bhattacharya2008,Hartmann2008,Heinrich2011,
	Massel2012,Xuereb2012,Xu2013,Liao2014,Nielsen2017,Piergentili2018,Yang2020}. Such systems enable the study of macroscopic collective quantum effects \cite{Chegnizadeh2024} and collective mechanical quantum-state engineering \cite{Xu2013,Schmidt2012,Buchmann2015}. Recently, the generation of macroscopic mechanical entanglement has been demonstrated in multimode optomechanical systems \cite{OckeloenKorppi2018,Kotler2021}. In particular, a common cavity field can mediate mechanical interactions \cite{Shkarin2014,Spethmann2016} and coherent excitation mixing and transport \cite{Lin2010,Mathew2020}. The shared cavity allows quantum interference among the coupling pathways, giving rise to collective mechanical modes with different couplings to the cavity.
	
	A characteristic consequence of quantum interference in the multiple-mechanical-mode optomechanical systems is the formation of mechanical dark modes \cite{Genes2008,OckeloenKorppi2019,Huang2023DarkMode}. For frequency-matched mechanical modes with beam-splitter couplings to a common cavity, destructive interference produces collective mechanical superpositions that decouple from the cavity field \cite{Massel2012,Shkarin2014}. The resulting suppression of the cavity-induced damping can protect collective mechanical excitations \cite{Shkarin2014}. More generally, dark-mode interference can reduce the influence of dissipation on excitation transfer \cite{Wang2012,Tian2012,WangClerk2012NJP,Dong2012}. For mechanical dark modes, however, decoupling also prevents the extraction of the thermal excitations through the cavity. Therefore, the dark-mode effect will hinder simultaneous cooling \cite{Genes2008,OckeloenKorppi2019,Lai2020,Lai2021,Huang2022a,Liu2022,Cao2025,Piotrowski2023,Xu2024} and limit optomechanical entanglement \cite{Huang2022b,Lai2022} and mechanical squeezing \cite{Huang2023} at finite temperatures. 
	
	To control the excitation transfer in multimode optomechanical systems, it is needed to manipulate the dark-mode effect by developing suitable active control schemes. So far, several methods have been proposed to control the dark modes by modifying either the coupling structure or the interference conditions. For example, the auxiliary cavity method has been proposed to break the dark-mode effect by introducing additional interaction and cooling channels \cite{Huang2022a,Liu2022,Cao2025,Huang2022b}. Alternatively, a phase-dependent mechanical coupling has been suggested to break the dark-mode effect by controlling the interference in a closed optomechanical interaction loop \cite{LaiOMIT2020,Lai2022,Huang2023}. Such phase-dependent coupling can also induce nonreciprocal phonon transfer and enable simultaneous cooling of mechanical modes \cite{Lai2020}. For two mechanical modes coupled to a common cavity	without direct mechanical coupling, a frequency mismatch mixes the bright and dark modes \cite{Lai2020}. This suggests using the effective mechanical frequency mismatch to control excitation transfer to the cavity.
	
	Motivated by the fact that periodic modulation offers a means for engineering effective frequencies and interactions \cite{Goldman2014,Bukov2015,Liao2016}, here, we propose a Floquet sideband method	to manipulate the dark-mode effect in a three-mode optomechanical system consisting of a cavity mode coupled to two mechanical modes. We periodically modulate one mechanical frequency to generate the Floquet sidebands and select a sideband to establish effective frequency matching with the other mechanical mode. We derive an approximate beam-splitter Hamiltonian supporting collective bright and dark modes under effective frequency matching. We determine the semiclassical operating point self-consistently and examine linear stability of the system within the rotating-wave approximation (RWA). The validity of the approximate Hamiltonian is assessed using the fidelity between the states evolved under the approximate Hamiltonian and the exact time-dependent Hamiltonian. The population trajectories show that a finite effective mechanical frequency mismatch couples the dark and bright modes, allowing excitation to dissipate through the cavity. Restoring frequency matching suppresses further transfer and protects the remaining dark-mode population. For the pulse duration and amplitudes considered, increasing the detuning amplitude reduces the dark-mode population remaining after the pulse.

	The remainder of this paper is organized as follows. In Sec.~\ref{sec:2}, we introduce the three-mode optomechanical model and its Hamiltonian. In Sec.~\ref{sec:3}, we derive the linearized optomechanical Hamiltonian associated with the selected Floquet sideband. In Sec.~\ref{sec:4}, we determine the semiclassical steady states and examine the linear stability of the system. In Sec.~\ref{sec:5}, we assess the validity of the approximate Hamiltonian by examining the fidelity between the two states governed by the approximate and exact Hamiltonians. In Sec.~\ref{sec:6}, we evaluate the dark-mode manipulation by analyzing the population dynamics of the system. Finally, we conclude this work in Sec.~\ref{sec:7}.
	
	\section{PHYSICAL MODEL AND HAMILTONIANS\label{sec:2}}
	\begin{figure}[ht]
		\centering
		\includegraphics[width=0.48\textwidth]{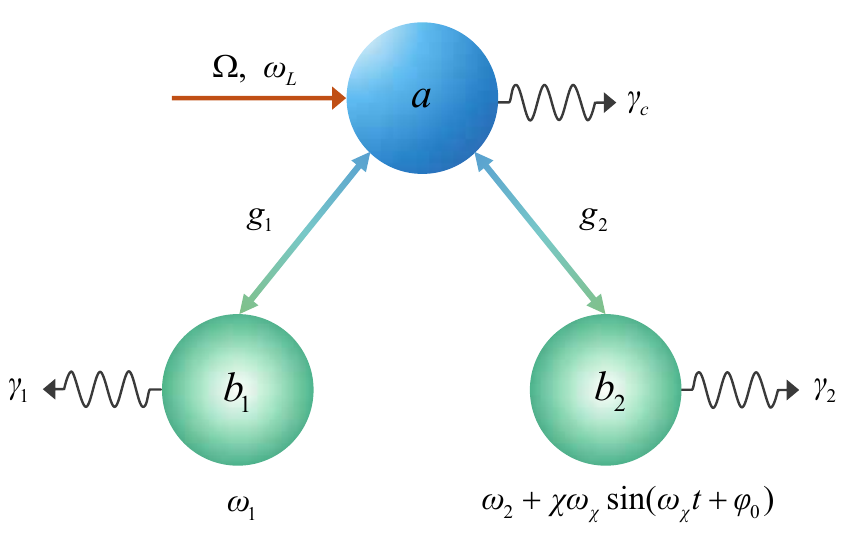}
		\caption{\label{fig:1}The system consists of a cavity mode $a$ (frequency $\omega_c$) and two mechanical modes $b_1$ and $b_2$ (frequencies $\omega_1$ and $\omega_2$, decay rates $\gamma_{1}$ and $\gamma_{2}$). The cavity field is driven by a coherent field  with amplitude $\Omega$ and frequency $\omega_L$. The coupling strengths between the cavity and the two mechanical modes are $g_1$ and $g_2$. The second mechanical mode $b_2$ has the time-dependent frequency $\omega_2+\chi \omega _{\chi}\sin(\omega _{\chi}t+\varphi _{0})$.}
	\end{figure}
	We consider a three-mode optomechanical system, which consists of a cavity mode coupled to two mechanical modes. Here, the cavity mode is characterized by the annihilation (creation) operator $\hat{a}$ ($\hat{a}^\dagger$) and frequency $\omega_c$, while the two mechanical modes are described by the annihilation (creation) operators $\hat{b}_1$ and $\hat{b}_2$ ($\hat{b}_1^\dagger$ and $\hat{b}_2^\dagger$), as well as the corresponding frequencies $\omega_{1}$ and $\omega_{2}$ \cite{Kippenberg2008,Aspelmeyer2014}. The total Hamiltonian of the system reads ($\hbar=1$)
	\begin{align}\label{Eq1}
		\hat{H} &= \omega _{c}\hat{a}^{\dagger}\hat{a}+\omega _{1}\hat{b}_{1}^{\dag}\hat{b}_{1}+\omega_{2}\hat{b}_{2}^{\dag }\hat{b}_{2}-g_{1}\hat{a}^{\dagger}\hat{a}(\hat{b}_{1}^{\dag }+\hat{b}_{1})\notag \\
		&\quad-g_{2}\hat{a}^{\dagger }\hat{a}( \hat{b}_{2}^{\dag }+\hat{b}_{2})	+\Omega (\hat{a}^{\dagger }e^{-i\omega_{L}t}+\hat{a}e^{i\omega_{L}t})\notag \\
		&\quad+\chi \omega _{\chi}\sin(\omega _{\chi}t+\varphi _{0}) \hat{b}_{2}^{\dag}\hat{b}_{2}.
	\end{align}
	Here, $g_1$ and $g_2$ denote the single-photon optomechanical coupling  strengths, $\Omega$ is the amplitude of the external laser drive at frequency $\omega_{L}$, and the last term describes a parametric modulation of the second mechanical mode, characterized by the dimensionless modulation depth $\chi$, frequency $\omega_{\chi}$, and phase $\varphi_{0}$.
	
	The quantum systems are inevitably coupled to their environments. In our system, we consider the case where the cavity field and the two mechanical modes are coupled to a vacuum bath and  two independent thermal baths, respectively. In this case, the dynamics of the system is governed by the quantum master equation,
	\begin{align}\label{Eq2}
		\dot{\hat{\rho}}&=i[\hat{\rho},\hat{H}] +\frac{\gamma _{c}}{2}\mathcal{L}_{\hat{a}}[\hat{\rho}]\notag\\
		&\quad+\sum_{j=1}^2 \frac{\gamma_{j}}{2}[(\bar{n}_{\mathrm{th},j}+1) \mathcal{L}_{\hat{b}_j}[\hat{\rho}]+\bar{n}_{\mathrm{th},j} \mathcal{L}_{{\hat{b}}_j^{\dagger}}[\hat{\rho}]],
	\end{align}
	where $\hat{\rho}$ is the density matrix of the three-mode optomechanical system, and $\hat{H}$ is the system Hamiltonian given in Eq.~\eqref{Eq1}. In Eq.~\eqref{Eq2}, the Lindblad superoperator for a jump operators $\hat{O}$ ($\hat{O}=\hat{a},\hat{b}_1,\hat{b}_2$) is defined by
	\begin{align}\label{Eq3}
		\mathcal{L}_{\hat{O}}[\hat{\rho}]=2\hat{O} \hat{\rho} \hat{O}^\dagger-\hat{O}^\dagger \hat{O}\hat{\rho}-\hat{\rho} \hat{O}^\dagger \hat{O}.
	\end{align}
	Here, $\gamma_{c}$ and $\gamma_{j}$ ($j=1,2$) denote the decay rates of the cavity field and the mechanical modes, respectively, while $\bar{n}_{\mathrm{th},j}$ denotes the mean thermal phonon number of the $j$th mechanical bath. The Lindblad formulation guarantees that the quantum master equation preserves both the trace and positivity of the density matrix of the system.
	
	To eliminate the explicit time dependence arising from the laser drive term in the quantum master equation, we transform into a frame rotating at the laser frequency $\omega_L$. To this end, we introduce the unitary operator
	\begin{align}\label{Eq4}
		\hat{R}(t) =e^{-i\omega _{L}t\hat{a}^{\dagger}\hat{a}},
	\end{align}
	and perform the transformation
	\begin{align}\label{Eq5}
		\hat{\rho} =\hat{R}(t)\hat{\tilde{\rho}}\hat{R}^{\dagger}(t),
	\end{align}
	where $\hat{\tilde{\rho}}$ is the density matrix of the system in the transformed representation. Under this transformation, the cavity field operators acquire phase factors: $\hat{R}^{\dagger}(t) \hat{a}\hat{R}(t)=\hat{a}e^{-i\omega_{L}t}$ and $\hat{R}^{\dagger}(t)\hat{a}^{\dagger}\hat{R}(t)=\hat{a}^{\dagger}e^{i\omega_{L}t}$, while the mechanical oscillator operators $\hat{b}_1$ and $\hat{b}_2$ are not affected by this transformation. After performing the transformation, the quantum master equation \eqref{Eq2} becomes
	\begin{align}\label{Eq6}
		\dot{\hat{\tilde{\rho}}}&=i[\hat{\tilde{\rho}},\hat{H}_{\mathrm{rot}}] +\frac{\gamma _{c}}{2}\mathcal{L}_{\hat{a}}[\hat{\tilde{\rho}}]\notag\\
		&\quad+\sum_{j=1}^2 \frac{\gamma_{j}}{2}[(\bar{n}_{\mathrm{th},j}+1) \mathcal{L}_{\hat{b}_j}[\hat{\tilde{\rho}}]+\bar{n}_{\mathrm{th},j} \mathcal{L}_{{\hat{b}}_j^{\dagger}}[\hat{\tilde{\rho}}]],
	\end{align}
	where the transformed Hamiltonian in the rotating frame is given by
	\begin{align}\label{Eq7}
		\hat{H}_{\mathrm{rot}}&=\Delta_{c}\hat{a}^{\dagger}\hat{a}+\omega_{1}\hat{b}_{1}^{\dagger}\hat{b}_{1}+\omega_{2}\hat{b}_{2}^{\dagger}\hat{b}_{2}-g_{1}\hat{a}^{\dagger}\hat{a} (\hat{b}_{1}^{\dagger}+\hat{b}_{1}) \notag\\
		&\quad-g_{2}\hat{a}^{\dagger}\hat{a}(\hat{b}_{2}^{\dagger}+\hat{b}_{2})+\Omega(\hat{a}^{\dagger}+\hat{a}) \notag\\
		&\quad+\chi \omega_{\chi}\sin(\omega_{\chi}t+\varphi_{0})\hat{b}_{2}^{\dagger}\hat{b}_{2},
	\end{align}
	with $\Delta _{c}=\omega _{c}-\omega_{L}$ being the cavity field detuning. The dissipative terms retain their original form, as the transformation is unitary and commutes with the Lindblad operators for the mechanical modes. Moreover, for the cavity, the form of $\mathcal{L}_{\hat{a}}$ remains invariant under the phase rotation of $\hat{a}$. For notational simplicity, we henceforth write $\hat{\tilde{\rho}}$ as $\hat{\rho}$.
	
	\section{Linearized optomechanical Hamiltonian associated with the selected Floquet sideband \label{sec:3}}
	To describe the optomechanical interactions associated with the selected Floquet sideband, we derive an approximate Hamiltonian. We begin with a displacement transformation to eliminate the linear cavity-field terms arising from the drive $\Omega$. The displacement operators are defined as
	\begin{subequations}\label{Eq8}
		\begin{align}
			\hat{D}_{a}(\alpha) &= \exp(\alpha \hat{a}^{\dagger}-\alpha^{\ast}\hat{a}), \label{Eq8a}\\
			\hat{D}_{b_{1}}(\beta_{1}) &= \exp(\beta_{1}\hat{b}_{1}^{\dagger}-\beta_{1}^{\ast}\hat{b}_{1}),  \label{Eq8b}\\
			\hat{D}_{b_{2}}(\beta_{2}) &= \exp(\beta_{2}\hat{b}_{2}^{\dagger}-\beta_{2}^{\ast}\hat{b}_{2}). \label{Eq8c}
		\end{align}
	\end{subequations}
    The displacement operators transform the annihilation operators according to $\hat{D}_{a}^{\dagger}(\alpha)\hat{a}\hat{D}_{a}(\alpha)=\hat{a}+\alpha$, $\hat{D}_{b_{1}}^{\dagger}(\beta_{1})\hat{b}_{1}\hat{D}_{b_{1}}(\beta_{1})=\hat{b}_{1}+\beta_{1}$, and $\hat{D}_{b_{2}}^{\dagger}(\beta_{2})\hat{b}_{2}\hat{D}_{b_{2}}(\beta_{2})=\hat{b}_{2}+\beta_{2}$. The transformed density matrix is defined as
	\begin{align}\label{Eq9}
		\hat{\rho}=\hat{D}_{a}^{\dagger}(\alpha)\hat{D}_{b_{1}}^{\dagger}(\beta_{1})\hat{D}_{b_{2}}^{\dagger}(\beta_{2})\hat{\rho}^{\prime}\hat{D}_{a}(\alpha)\hat{D}_{b_{1}}(\beta_{1})\hat{D}_{b_{2}}(\beta_{2}),
	\end{align}
	where $\hat{\rho}^{\prime}$ is the density matrix of the system in the displaced representation. The displacement amplitudes $\alpha\left(t\right)$ and $\beta_{j}\left(t\right)$ are chosen to cancel the linear terms in the displaced Hamiltonian. 
	
	Inserting this transformation into the quantum master equation \eqref{Eq6} and using the properties of the displacement operators, we obtain the following relations,
	\begin{subequations}\label{Eq10}
		\begin{align}
			\dot{\alpha}&=(-i\Delta_{c}^{\prime}-\gamma_{c}/ 2)\alpha+i\Omega, \label{Eq10a}\\
			\dot{\beta_{1}}&=(-i\omega_{1}-\gamma_{1}/2)\beta_{1}-ig_{1}\left|\alpha\right|^{2}, \label{Eq10b}\\
			\dot{\beta_{2}}&=\{-i\left[\omega_{2}+\chi \omega_{\chi}\sin (\omega_{\chi}t+\varphi_{0}) \right]-\gamma_{2}/2\}\beta_{2}\notag\\
			&\quad-ig_{2}\left|\alpha\right|^{2}, \label{Eq10c}
		\end{align}
	\end{subequations}
	where the effective cavity detuning is defined as 
	\begin{align}
			\Delta_{c}^{\prime}=\Delta_{c}+g_{1}( \beta_{1}^{\ast}+\beta_{1}) +g_{2}(\beta_{2}^{\ast}+\beta_{2}).\label{Eq11}
	\end{align}
	Then the quantum master equation becomes
	\begin{align}\label{Eq12}
		\dot{\hat{\rho}}^{\prime}&=i[\hat{\rho}^{\prime},\hat{H}_{\mathrm{disp}}] +\frac{\gamma_{c}}{2}\mathcal{L}_{\hat{a}}[\hat{\rho}^{\prime}]\notag\\
		&\quad+\sum_{j=1}^2 \frac{\gamma_{j}}{2}[(\bar{n}_{\mathrm{th},j}+1) \mathcal{L}_{\hat{b}_j}[\hat{\rho}^{\prime}]+\bar{n}_{\mathrm{th},j} \mathcal{L}_{{\hat{b}}_j^{\dagger}}[\hat{\rho}^{\prime}]],
	\end{align}
	with the displaced Hamiltonian
	\begin{align}\label{Eq13}
		\hat{H}_{\mathrm{disp}}&=\Delta _{c}^{\prime}\hat{a}^{\dagger}\hat{a}+\omega_{1}\hat{b}_{1}^{\dagger}\hat{b}_{1}+[\omega_{2}+\chi\omega_{\chi}\sin(\omega_{\chi}t+\varphi_{0})]\hat{b}_{2}^{\dagger}\hat{b}_{2} \notag\\
		&\quad-g_{1}\hat{a}^{\dagger}\hat{a}(\hat{b}_{1}^{\dagger}+\hat{b}_{1})-g_{2}\hat{a}^{\dagger}\hat{a}(\hat{b}_{2}^{\dagger}+\hat{b}_{2}) \notag \\
		&\quad+g_{1}(\alpha \hat{a}^{\dagger}+\alpha^{\ast}\hat{a})(\hat{b}_{1}^{\dagger}+\hat{b}_{1}) \notag\\
		&\quad+g_{2}(\alpha \hat{a}^{\dagger}+\alpha^{\ast}\hat{a})(\hat{b}_{2}^{\dagger}+\hat{b}_{2}).
	\end{align}
	Here, $\hat a$ and $\hat b_j$ denote fluctuation operators in the displaced frame.
	
	The Hamiltonian $\hat{H}_{\mathrm{disp}}$ still contains explicit time dependence through the modulation term $\chi\omega_{\chi}\sin(\omega_{\chi}t+\varphi_{0})$. To remove this diagonal time dependence, we introduce a second unitary transformation. We define $\hat{H}_{0}(t) =\Delta_{c}^{\prime}\hat{a}^{\dagger}\hat{a}	+\omega_{1}\hat{b}_{1}^{\dagger}\hat{b}_{1}	+[\omega_{2}+\chi\omega_{\chi}\sin(\omega_{\chi}t+\varphi_{0})]
	\hat{b}_{2}^{\dagger}\hat{b}_{2}$. Since all terms are diagonal in the number operators, we have $[\hat{H}_{0}(t_{1}),\hat{H}_{0}(t_{2})] = 0$ for any $t_1$, $t_2$, and the time-ordered exponential reduces to an ordinary exponential, allowing us to define
	\begin{align}\label{Eq14}
		\hat{V}(t) = \mathcal{T}\!\exp\!\Big[ -i\int_{0}^{t}\hat{H}_{0}(\tau)\,d\tau \Big]
		= \exp\!\Big[ -i\int_{0}^{t}\hat{H}_{0}(\tau)\,d\tau \Big],
	\end{align}
	where $\mathcal{T}$ is the time-ordering operator. Evaluating the integral, we obtain $\hat{V}(t) = \exp\{-i\Delta_{c}^{\prime}t\,\hat{a}^{\dagger}\hat{a}	- i\omega_{1}t\,\hat{b}_{1}^{\dagger}\hat{b}_{1}
	- i[\omega_{2}t - \chi\cos(\omega_{\chi}t+\varphi_{0})+\chi\cos \varphi_{0}]\hat{b}_{2}^{\dagger}\hat{b}_{2}\}$. Under this transformation, the operators evolve as	$\hat{V}^{\dagger}(t)\,\hat{a}\,\hat{V}(t) = \hat{a}\,e^{-i\Delta_{c}^{\prime}t}$, $\hat{V}^{\dagger}(t)\,\hat{b}_{1}\,\hat{V}(t) = \hat{b}_{1}\,e^{-i\omega_{1}t}$, and $\hat{V}^{\dagger}(t)\,\hat{b}_{2}\,\hat{V}(t) = \hat{b}_{2}\,e^{- i\omega_{2}t}\,e^{i\chi [\cos (\omega_{\chi}t + \varphi_{0})-\cos\varphi_{0}]}$. We now transform the density matrix as
	\begin{align}\label{Eq15}
		\hat{\rho}^{\prime\prime}(t) = \hat{V}^{\dagger}(t)\,\hat{\rho}^{\prime}\,\hat{V}(t),
	\end{align}
	and then the transformed quantum master equation becomes
	\begin{align}\label{Eq16}
		\dot{\hat{\rho}}^{\prime\prime} &= i[\hat{\rho}^{\prime\prime},\hat{H}_{1}(t)] + \frac{\gamma_{c}}{2}\mathcal{L}_{\hat{a}}[\hat{\rho}^{\prime\prime}] \notag\\
		&\quad + \sum_{j=1}^2 \frac{\gamma_{j}}{2}\big[(\bar{n}_{\mathrm{th},j}+1) \mathcal{L}_{\hat{b}_j}[\hat{\rho}^{\prime\prime}]
		+ \bar{n}_{\mathrm{th},j}\,\mathcal{L}_{\hat{b}_j^{\dagger}}[\hat{\rho}^{\prime\prime}]\big],
	\end{align}
	where the Hamiltonian $\hat{H}_1(t)$ is introduced as
	\begin{align}\label{Eq17}
		\hat{H}_1(t)&=\hat{V}^{\dagger}(t)\,\hat{H}_{\mathrm{disp}}\,\hat{V}(t) - \hat{H}_{0}(t)\notag\\
		&=\bigl[ -g_1 \hat{a}^\dagger \hat{a} \hat{b}_1^\dagger e^{i\omega_1 t}\notag\\
		&\quad	- g_2 \hat{a}^\dagger \hat{a} \hat{b}_2^\dagger e^{i\omega_2 t}e^{-i\chi [\cos (\omega_{\chi}t + \varphi_{0})-\cos\varphi_{0}]} \notag\\
		&\quad + g_1 \alpha \hat{a}^\dagger \hat{b}_1^\dagger e^{i(\Delta_c' + \omega_1)t} + g_1 \alpha \hat{a}^\dagger \hat{b}_1 e^{i(\Delta_c' - \omega_1)t} \notag\\
		&\quad + g_2 \alpha \hat{a}^\dagger \hat{b}_2^\dagger e^{i(\Delta_c' + \omega_2)t} e^{-i\chi [\cos (\omega_{\chi}t + \varphi_{0})-\cos\varphi_{0}]} \notag\\
		&\quad + g_2 \alpha \hat{a}^\dagger \hat{b}_2 e^{i(\Delta_c' - \omega_2)t} e^{i\chi [\cos (\omega_{\chi}t + \varphi_{0})-\cos\varphi_{0}]} \bigr] + \mathrm{H.c.}
	\end{align}
	To proceed, we expand the exponentials containing $\cos(\omega_{\chi}t+\varphi_{0})$ using the Jacobi--Anger identity:
	\begin{align}\label{Eq18}
		e^{\pm i\chi\cos(\omega_{\chi}t+\varphi_{0})} =	\overset{\infty}{\underset{n=-\infty}{\sum}} (\pm i)^{n}J_{n}(\chi)\, e^{in(\omega_{\chi}t+\varphi_{0})},
	\end{align}
	where $J_{n}(\chi)$ is the Bessel function of the first kind of integer order $n$. Substituting these expansions into $\hat{H}_{1}(t)$ yields
	\begin{align}\label{Eq19}
		\hat{H}_{1}(t) &= -g_{1}\hat{a}^{\dagger}\hat{a}\big(\hat{b}_{1}^{\dagger}e^{i\omega_{1}t} + \hat{b}_{1}e^{-i\omega_{1}t}\big)-g_{2}\hat{a}^{\dagger}\hat{a} \notag \\
		&\quad \times \overset{\infty}{\underset{n=-\infty}{\sum}} \bigl[(-i)^{n}J_{n}(\chi)e^{-in\varphi_{0}}e^{i\chi\cos\varphi_{0}}\hat{b}_{2}^{\dagger}e^{i(\omega_{2}-n\omega_{\chi})t} \notag\\
		&\quad + i^{n}J_{n}(\chi)e^{in\varphi_{0}}e^{-i\chi\cos\varphi_{0}}\hat{b}_{2}e^{-i(\omega_{2}-n\omega_{\chi})t} \bigr] \notag\\
		&\quad + g_{1}(\alpha \hat{a}^{\dagger}e^{i\Delta_{c}^{\prime}t} + \alpha^{\ast}\hat{a}e^{-i\Delta_{c}^{\prime}t}) (\hat{b}_{1}^{\dagger}e^{i\omega_{1}t} + \hat{b}_{1}e^{-i\omega_{1}t}) \notag \\
		&\quad + g_{2}(\alpha \hat{a}^{\dagger}e^{i\Delta_{c}^{\prime}t} + \alpha^{\ast}\hat{a}e^{-i\Delta_{c}^{\prime}t}) \notag \\
		&\quad \times \overset{\infty}{\underset{n=-\infty}{\sum}} \bigl[(-i)^{n}J_{n}(\chi)e^{-in\varphi_{0}}e^{i\chi\cos\varphi_{0}}\hat{b}_{2}^{\dagger}e^{i(\omega_{2}-n\omega_{\chi})t} \notag\\
		&\quad + i^{n}J_{n}(\chi)e^{in\varphi_{0}}e^{-i\chi \cos\varphi_{0}}\hat{b}_{2}e^{-i(\omega_{2}-n\omega_{\chi})t}\bigr].
	\end{align}
	The summations over $n$ in Eq.~\eqref{Eq19} represent the Floquet sidebands generated by the periodic modulation, with $n$ labeling the sideband order and $J_n(\chi)$ determining the corresponding modulation-dependent weight. We now select a specific sideband by choosing an integer $n_0$ such that
	\begin{align}\label{Eq20}
		\omega_{2} - n_{0}\omega_{\chi} = \delta_{2}.
	\end{align}
	For low occupations of the fluctuation modes, the nonlinear radiation-pressure terms are neglected when $\omega_1\gg|g_1|$ and		$|\omega_2-n\omega_\chi|\gg|g_2J_n(\chi)|$.	The counter-rotating terms are neglected under the conditions $|\Delta_c'+\omega_1|\gg|g_1\alpha|$ and	$|\Delta_c'+\omega_2-n\omega_\chi|\gg|g_2\alpha J_n(\chi)|$. For the beam-splitter terms, the detuning of sideband	$n$ is $\Delta_c'-\delta_2+(n-n_0)\omega_\chi$. Thus, the nonselected sidebands can be neglected when $|\Delta_c'-\delta_2|\ll\omega_\chi$ and $|(n-n_0)\omega_\chi|\gg|g_2\alpha J_n(\chi)|$ for $n\neq n_0$. Retaining the beam-splitter interactions with the first mechanical mode and the selected Floquet sideband $n=n_0$, we obtain the approximate Hamiltonian
	\begin{align}\label{Eq21}
		\hat{H}_{\mathrm{app}}(t)& = g_{1}\bigl[\alpha \hat{a}^{\dagger}\hat{b}_{1}e^{i(\Delta_{c}^{\prime} -\omega_{1})t} + \alpha^{\ast}\hat{a}\hat{b}_{1}^{\dagger}e^{-i(\Delta_{c}^{\prime}-\omega_{1})t}\bigr] \notag \\
		&\quad + g_{2}i^{n_{0}}J_{n_{0}}(\chi)e^{in_{0}\varphi_{0}}e^{-i\chi \cos\varphi_{0}}\alpha\hat{a}^{\dagger}\hat{b}_{2}	e^{i(\Delta_{c}^{\prime}-\delta_{2})t}\notag\\
		&\quad + g_2(-i)^{n_{0}}J_{n_{0}}(\chi)e^{-in_{0}\varphi_{0}}e^{i\chi \cos\varphi_{0}}\notag\\
		&\qquad\times\alpha^{\ast}\hat{a}\hat{b}_{2}^{\dagger}	e^{-i(\Delta_{c}^{\prime}-\delta_{2})t} .
	\end{align}	
	For convenience, we define the effective coupling strengths $g_{\mathrm{1,eff}} = g_{1}\alpha$, and $g_{\mathrm{2,eff}} = g_{2}J_{n_{0}}(\chi)e^{in_{0}({\pi}/{2}+\varphi_{0})}e^{-i\chi \cos\varphi_{0}}\alpha$, where we have used $i^{n_{0}} = e^{{in_{0}\pi}/{2}}$, and $(-i)^{n_{0}} = e^{{-in_{0}\pi}/{2}}$. Then
	\begin{align}\label{Eq22}
		\hat{H}_{\mathrm{app}}(t) &= g_{\mathrm{1,eff}}\,\hat{a}^{\dagger}\hat{b}_{1}e^{i(\Delta_{c}^{\prime}-\omega_{1})t}	+ g_{\mathrm{1,eff}}^{\ast}\,\hat{a}\hat{b}_{1}^{\dagger}e^{-i(\Delta_{c}^{\prime}-\omega_{1})t} \notag \\
		&\quad + g_{\mathrm{2,eff}}\,\hat{a}^{\dagger}\hat{b}_{2}e^{i(\Delta_{c}^{\prime}-\delta_{2})t}	+ g_{\mathrm{2,eff}}^{\ast}\,\hat{a}\hat{b}_{2}^{\dagger}e^{-i(\Delta_{c}^{\prime}-\delta_{2})t}.
	\end{align}
	To eliminate the residual time dependence and obtain a fully time-independent Hamiltonian, we further apply the unitary transformation
	\begin{align}\label{Eq23}
		\hat{U}(t) = \exp\!\big[ -i(\Delta_{c}^{\prime}-\omega_{1})\hat{b}_{1}^{\dagger}\hat{b}_{1}t - i(\Delta_{c}^{\prime}-\delta_{2})\hat{b}_{2}^{\dagger}\hat{b}_{2}t \big].
	\end{align}
	Under this transformation, the Hamiltonian in Eq.~\eqref{Eq22} becomes
	\begin{align}\label{Eq24}
		\hat{H}_{\mathrm{app}} &= g_{\mathrm{1,eff}}\,\hat{a}^{\dagger}\hat{b}_{1} + g_{\mathrm{1,eff}}^{\ast}\,\hat{a}\hat{b}_{1}^{\dagger} + g_{\mathrm{2,eff}}\,\hat{a}^{\dagger}\hat{b}_{2}
		+ g_{\mathrm{2,eff}}^{\ast}\,\hat{a}\hat{b}_{2}^{\dagger} \notag\\
		&\quad - \bigl[(\Delta_{c}^{\prime}-\omega_{1})\hat{b}_{1}^{\dagger}\hat{b}_{1}	+ (\Delta_{c}^{\prime}-\delta_{2})\hat{b}_{2}^{\dagger}\hat{b}_{2}\bigr].
	\end{align}
	The residual detunings now appear entirely in the diagonal mechanical terms. Under the single-photon resonance condition
	\begin{align}\label{Eq25}
		\Delta_{c}^{\prime} - \omega_{1} = \Delta_{c}^{\prime} - \delta_{2} = 0,
	\end{align}
	i.e., $\Delta_{c}^{\prime} = \omega_{1} = \delta_{2}$, the Hamiltonian is reduced to the beam-splitter form
	\begin{align}\label{Eq26}
		\hat{H}_{\mathrm{app}}^{\prime} &= g_{\mathrm{1,eff}}\,\hat{a}^{\dagger}\hat{b}_{1}	+ g_{\mathrm{1,eff}}^{\ast}\,\hat{a}\hat{b}_{1}^{\dagger} + g_{\mathrm{2,eff}}\,\hat{a}^{\dagger}\hat{b}_{2} 
		+ g_{\mathrm{2,eff}}^{\ast}\,\hat{a}\hat{b}_{2}^{\dagger}.
	\end{align}
	This Hamiltonian describes a coherent excitation exchange between the cavity and each mechanical mode. Equation~\eqref{Eq26} is written in the fully transformed frame defined by the successive application of $\hat{V}(t)$ and $\hat{U}(t)$, where all explicit time dependence has been eliminated and $\hat{b}_2$ evolves at the effective frequency $\delta_2 = \omega_2 - n_0\omega_\chi$. This beam-splitter Hamiltonian provides the basis for the dark-mode analysis presented in Sec.~\ref{sec:6}.
	
	\section{Semiclassical motion and Stability Analysis\label{sec:4}}
	
	\subsection{Semiclassical motion of the system}	\label{sec:4A}
	
	The effective coupling strengths derived in Sec.~\ref{sec:3} depend on the classical cavity displacement and must therefore be evaluated at a self-consistent semiclassical operating point. Because the effective cavity detuning depends on the mechanical displacements, the semiclassical equations are nonlinear and may admit multiple solutions. We first derive the reduced equations for the coupled semiclassical amplitudes by retaining both the selected Floquet sideband and the slow mechanical feedback to the cavity. Their stationary solutions can then be determined self-consistently.
	
	We first account for the periodic frequency modulation of the second mechanical mode by introducing the transformation
	\begin{align}
		\beta_2(t)=e^{-i\omega_2t}e^{i\chi\cos(\omega_\chi t+\varphi_0)} \bar{\beta}_2(t). \label{Eq27}
	\end{align}
	This transformation removes the free oscillation at $\omega_2$ and the diagonal modulation from Eq.~\eqref{Eq10c}, which then takes the form
	\begin{align}
		\dot{\bar{\beta}}_2	= -\frac{\gamma_2}{2}\bar{\beta}_2 -i g_2|\alpha|^2e^{i\omega_2t} e^{-i\chi\cos(\omega_\chi t+\varphi_0)}. \label{Eq28}
	\end{align}
	Using the Jacobi--Anger expansion in Eq.~\eqref{Eq18}, we obtain
	\begin{align}
		\dot{\bar{\beta}}_2	&= -\frac{\gamma_2}{2}\bar{\beta}_2	\notag\\
		&\quad-i g_2|\alpha|^2 \sum_{n=-\infty}^{\infty}	(-i)^nJ_n(\chi)e^{-in\varphi_0}	e^{i(\omega_2-n\omega_\chi)t}.	\label{Eq29}
	\end{align}
	The summation in Eq.~\eqref{Eq29} corresponds to the Floquet sidebands generated by the periodic modulation of the mechanical frequency. Following the derivation of the approximate Hamiltonian, we retain the sideband $n=n_0$ selected in Eq.~\eqref{Eq20}. The terms with $n\neq n_0$ oscillate rapidly relative to the selected-sideband dynamics and are neglected under the rotating-wave approximation.
	Equation~\eqref{Eq29} is therefore reduced to
	\begin{align}
		\dot{\bar{\beta}}_2	\simeq -\frac{\gamma_2}{2}\bar{\beta}_2 -i g_2(-i)^{n_0}J_{n_0}(\chi)e^{-in_0\varphi_0}|\alpha|^2e^{i\delta_2t}.\label{Eq30}
	\end{align}
	To remove the phase factor $e^{i\delta_2t}$ from the driving term, we introduce the selected-sideband amplitude
	\begin{align}
		\bar{\beta}_2(t) = e^{i\delta_2t}\beta_2^{[n_0]}(t). \label{Eq31}
	\end{align}
	Substitution of Eq.~\eqref{Eq31} into Eq.~\eqref{Eq30} gives
	\begin{align}
		\dot{\beta}_2^{[n_0]} &= -\left(i\delta_2+\frac{\gamma_2}{2}\right)\beta_2^{[n_0]}-	i g_2(-i)^{n_0}J_{n_0}(\chi)e^{-in_0\varphi_0} |\alpha|^2. \label{Eq32}
	\end{align}
	Based on Eqs.~\eqref{Eq27} and \eqref{Eq31}, we can obtain the relation
	\begin{align}
		\beta_2(t) \simeq e^{-in_0\omega_\chi t} e^{i\chi\cos(\omega_\chi t+\varphi_0)}	\beta_2^{[n_0]}(t).	\label{Eq33}
	\end{align}
	Using the Jacobi--Anger expansion and discarding the rapidly oscillating terms, we can introduce a slow variable as
	\begin{align}
		\beta_2^{\mathrm{slow}}(t) =i^{n_0}J_{n_0}(\chi)e^{in_0\varphi_0}\beta_2^{[n_0]}(t). \label{Eq34}
	\end{align}
	Hereafter, we set $\varphi_0=0$. Differentiating Eq.~\eqref{Eq34} with respect to time and utilizing Eq.~\eqref{Eq32}, we obtain
	\begin{align}
		\dot{\beta}_2^{\mathrm{slow}} = -\left(i\delta_2+\frac{\gamma_2}{2}\right)	\beta_2^{\mathrm{slow}}-i g_2J_{n_0}^{2}(\chi)|\alpha|^2. \label{Eq35}
	\end{align}
	
	Substituting the slow component $\beta_2^{\mathrm{slow}}(t)$ into Eq.~\eqref{Eq11}, we obtain the normalized cavity detuning as
	\begin{align}
		\Delta_{c,\mathrm{slow}}'(t) &=	\Delta_c+g_1\left[\beta_1(t)+\beta_1^*(t)\right] \notag\\
		&\quad+g_2\left[\beta_2^{\mathrm{slow}}(t) +\beta_2^{\mathrm{slow}*}(t)	\right]. \label{Eq36}
	\end{align}
	With this detuning, Eq.~\eqref{Eq10a} takes the form
	\begin{align}
		\dot{\alpha} = -\left( i\Delta_{c,\mathrm{slow}}'+\frac{\gamma_c}{2}\right)\alpha+i\Omega. \label{Eq37}
	\end{align}
	Based on the above discussion, we obtain Eqs.~\eqref{Eq37}, \eqref{Eq10b}, and \eqref{Eq35} to describe the coupled evolution of $\alpha$, $\beta_1$, and $\beta_2^{\mathrm{slow}}$ within the reduced semiclassical model. These equations contain no explicit time-dependent coefficients. 
	
	Their stationary solutions satisfy $\dot{\alpha}=\dot{\beta}_1=\dot{\beta}_2^{\mathrm{slow}}=0$. From Eqs.~\eqref{Eq10b} and \eqref{Eq35}, the mechanical stationary amplitudes are
	\begin{subequations}\label{Eq38}
		\begin{align}
			\beta_{1,\mathrm{ss}} &= \frac{-i g_1|\alpha_{\mathrm{ss}}|^2} {i\omega_1+\gamma_1/2}, \label{Eq38a}\\
			\beta_{2,\mathrm{ss}}^{\mathrm{slow}} &= \frac{-i g_2J_{n_0}^{2}(\chi)|\alpha_{\mathrm{ss}}|^2}	{i\delta_2+\gamma_2/2}.	\label{Eq38b}
		\end{align}
	\end{subequations}
	Substituting Eqs.~\eqref{Eq38a} and \eqref{Eq38b} into Eq.~\eqref{Eq36}, we obtain the stationary effective cavity detuning
	\begin{align}
		\bar{\Delta}_c'	=\Delta_c-C|\alpha_{\mathrm{ss}}|^2, \label{Eq39}
	\end{align}
	where the total optomechanical frequency-shift coefficient is given by
	\begin{align}
		C=\frac{2g_1^2\omega_1}{\omega_1^2+(\gamma_1/2)^2}+	\frac{2g_2^2J_{n_0}^2(\chi)\delta_2}{\delta_2^2+(\gamma_2/2)^2}. \label{Eq40}
	\end{align}
	The two terms describe the contributions of the first mechanical mode and the selected Floquet sideband of the second mechanical mode to the cavity-frequency shift. The stationary solution of Eq.~\eqref{Eq37} is
	\begin{align}
		\alpha_{\mathrm{ss}}=\frac{i\Omega}{i\bar{\Delta}_c'+\gamma_c/2}. \label{Eq41}
	\end{align}
	Substituting Eq.~\eqref{Eq39} into Eq.~\eqref{Eq41} and taking the modulus squared gives
	\begin{align}
		|\alpha_{\mathrm{ss}}|^2 = \frac{\Omega^2} {\left(\Delta_c-C|\alpha_{\mathrm{ss}}|^2 \right)^2+\left(\gamma_c/2\right)^2}. \label{Eq42}
	\end{align}
	Equivalently, the steady-state intracavity photon number satisfies
	\begin{align}
		&C^2\left(|\alpha_{\mathrm{ss}}|^2\right)^3	-2\Delta_c C \left(|\alpha_{\mathrm{ss}}|^2\right)^2 \notag\\ &\quad+\left[\Delta_c^2+\left(\frac{\gamma_c}{2}\right)^2\right]|\alpha_{\mathrm{ss}}|^2-\Omega^2=	0.	\label{Eq43}
	\end{align}
	Each nonnegative real root of Eq.~\eqref{Eq43} defines a candidate semiclassical operating point within the reduced description. A unique nonnegative root corresponds to a single-valued response, whereas multiple nonnegative roots indicate a multivalued response. Once the photon number is determined, the cavity and mechanical stationary amplitudes can be determined based on Eqs.~\eqref{Eq41} and \eqref{Eq38}.
	
	To assess the accuracy of the single-sideband approximation, we compare the reduced stationary solution with a numerically converged periodic steady state of the original semiclassical motion governed by Eqs.~\eqref{Eq10a}--\eqref{Eq10c}. At the operating point used in the subsequent calculations, numerical results indicate that the relative differences in the period-averaged cavity amplitude, intracavity photon number, and first-mechanical-mode amplitude are of order $10^{-4}$ or smaller. The absolute difference between the reduced effective cavity detuning and the period-averaged value obtained from the original equations is of order $10^{-4}\omega_1$. These comparisons support the validity of the reduced stationary solution for determining the effective coupling strengths at the selected operating point.	
	
	\subsection{Analysis of the stability of quantum motion}\label{sec:4B}
	
	To further study the steady state obtained in Sec.~\ref{sec:4A}, we examine the linear dynamics of quantum fluctuations around the semiclassical operating point. The fluctuation operators in the displaced frame are denoted by $\delta\hat a$, $\delta\hat b_1$, and $\delta\hat b_2$, where $\delta\hat b_2$ is defined in the rotating frame associated with the selected Floquet sideband $n=n_0$. For keeping consistency with the dark-mode analysis presented below, we use the same rotating-wave approximate Hamiltonian as that given in Eq.~\eqref{Eq24}.
	
	At a given semiclassical steady state, the effective optomechanical coupling strengths are given by
	\begin{align}
		g_{1,\mathrm{eff}} &= g_1\alpha_{\mathrm{ss}}, \notag\\
		g_{2,\mathrm{eff}} &= g_2 J_{n_0}(\chi) e^{i n_0(\varphi_0+\pi/2)}e^{-i\chi \cos\varphi_{0}}\alpha_{\mathrm{ss}}. \label{Eq44}
	\end{align}
	The quantum Langevin equations corresponding to the approximate Hamiltonian in Eq.~\eqref{Eq24} are
	\begin{subequations}\label{Eq45}
	\begin{align}
		\delta\dot{\hat a} =&-\frac{\gamma _{c}}{2}\delta {\hat a}-ig_{1,\mathrm{eff}}\delta{\hat b}_{1}-ig_{2,\mathrm{eff}}\delta {\hat{b}}_{2}+\sqrt{\gamma _{c}}\hat{a}_{\mathrm{in}},\label{Eq45a}\\
		\delta \dot{\hat{b}}_{1} =&[i( \bar{\Delta} _{c}^{\prime }-\omega _{1}) -\frac{\gamma _{1}}{2}]\delta \hat{b}_{1}-ig_{1,\mathrm{eff}}^{\ast }\delta {\hat a}+\sqrt{\gamma_{1}}\hat{b}_{1,\mathrm{in}},	\label{Eq45b}\\
		\delta \dot{\hat{b}}_{2} =&[i( \bar{\Delta} _{c}^{\prime }-\delta _{2}) -\frac{\gamma _{2}}{2}]\delta \hat{b}_{2}-ig_{2,\mathrm{eff}}^{\ast }\delta {\hat a}+\sqrt{\gamma_{2}}\hat{b}_{\mathrm{2,in}},	\label{Eq45b}
	\end{align}
    \end{subequations}
   where $\hat a_{\mathrm{in}}$ and $\hat b_{j,\mathrm{in}}$ are the input-noise operators for the cavity and the $j$th mechanical mode, respectively. For a vacuum cavity bath and independent thermal mechanical baths, the nonzero correlation functions are $\langle\hat a_{\mathrm{in}}(t) \hat a_{\mathrm{in}}^\dagger(t')\rangle=\delta(t-t')$, $\langle\hat b_{j,\mathrm{in}}(t) \hat b_{k,\mathrm{in}}^\dagger(t')\rangle =(\bar n_{\mathrm{th},j}+1)\delta_{jk}\delta(t-t')$, and $\langle\hat b_{j,\mathrm{in}}^\dagger(t) \hat b_{k,\mathrm{in}}(t')\rangle =\bar n_{\mathrm{th},j}\delta_{jk}\delta(t-t')$ for $j,k\in\{1,2\}$.
	
	To express these equations in real matrix form, we introduce the quadrature operators
	\begin{align}\label{Eq46}
		\hat{X}_o = \frac{\hat{o} + \hat{o}^\dagger}{\sqrt{2}}, \quad
		\hat{Y}_o = \frac{i(\hat{o}^\dagger - \hat{o} )}{\sqrt{2}},
	\end{align}
	for $\hat{o} \in \{\delta \hat{a}, \delta \hat{b}_1, \delta \hat{b}_2\}$. We then introduce the quadrature operator vector $\boldsymbol{u} = (\hat{X}_a, \hat{Y}_a, \hat{X}_{b_1}, \hat{Y}_{b_1}, \hat{X}_{b_2}, \hat{Y}_{b_2})^\mathrm{T}$ and the corresponding noise operator vector 
	\begin{align}
		\boldsymbol{\xi}(t)&=\left(\sqrt{\gamma_c}\hat X_{a,\mathrm{in}},\sqrt{\gamma_c}\hat Y_{a,\mathrm{in}},	\sqrt{\gamma_1}\hat X_{b_1,\mathrm{in}},\right. \notag\\
		&\left.	\hspace{2em}
		\sqrt{\gamma_1}\hat Y_{b_1,\mathrm{in}},\sqrt{\gamma_2}\hat X_{b_2,\mathrm{in}},\sqrt{\gamma_2}\hat Y_{b_2,\mathrm{in}}	\right)^{\mathrm{T}}. \label{Eq47}
	\end{align}
	The input-noise quadratures in Eq.~\eqref{Eq47} are defined by $\hat X_{o,\mathrm{in}}=(\hat o_{\mathrm{in}}+\hat o_{\mathrm{in}}^\dagger)/\sqrt{2}$ and $\hat Y_{o,\mathrm{in}}=i(\hat o_{\mathrm{in}}^\dagger-\hat o_{\mathrm{in}})/\sqrt{2}$ for $\hat o_{\mathrm{in}}\in \{\hat a_{\mathrm{in}},\hat b_{1,\mathrm{in}}, \hat b_{2,\mathrm{in}}\}$. The linearized dynamics can then be written as
	\begin{align}
		\dot{\boldsymbol{u}}=\mathbf{A}\boldsymbol{u}+\boldsymbol{\xi}(t), \label{Eq48}
	\end{align}
	where the drift matrix $\mathbf{A}$ is given by
	\begin{align}
		\mathbf{A} =
		\begin{pmatrix}
			-\dfrac{\gamma _{c}}{2} & 0 & \tilde{g}_1^{\mathrm{I}} & \tilde{g}_1^{\mathrm{R}}  & \tilde{g}_2^{\mathrm{I}} & \tilde{g}_2^{\mathrm{R}}  \\[6pt]
			0 & -\dfrac{\gamma _{c}}{2} & -\tilde{g}_1^{\mathrm{R}}  & \tilde{g}_1^{\mathrm{I}} & -\tilde{g}_2^{\mathrm{R}}  & \tilde{g}_2^{\mathrm{I}} \\[6pt]
			-\tilde{g}_1^{\mathrm{I}} & \tilde{g}_1^{\mathrm{R}}  & -\dfrac{\gamma _{1}}{2} & -d_1  & 0 & 0 \\[6pt]
			-\tilde{g}_1^{\mathrm{R}}  & -\tilde{g}_1^{\mathrm{I}} & d_1  & -%
			\dfrac{\gamma _{1}}{2} & 0 & 0 \\[6pt]
			-\tilde{g}_2^{\mathrm{I}} & \tilde{g}_2^{\mathrm{R}}  & 0 & 0 & -\dfrac{\gamma _{2}}{2} & -d_2 \\[6pt]
			-\tilde{g}_2^{\mathrm{R}}  & -\tilde{g}_2^{\mathrm{I}} & 0 & 0 & d_2  & -\dfrac{\gamma _{2}}{2}%
		\end{pmatrix},
		\label{Eq49}
	\end{align}
	Here, $d_1=\bar{\Delta}_c'-\omega_1$ and $d_2=\bar{\Delta}_c'-\delta_2$ denote the cavity--mechanical detunings, while $\tilde{g}_j^{\mathrm{R}} \equiv \mathrm{Re}(g_{j,\mathrm{eff}})$ and $\tilde{g}_j^{\mathrm{I}} \equiv \mathrm{Im}(g_{j,\mathrm{eff}})$, with $j=1,2$. The linear stability of the fluctuation dynamics within the effective model is determined by the eigenvalues $\lambda_\ell$ of the drift matrix $\mathbf{A}$. The stability condition is given by
	\begin{align}
		\max_{\ell}\left\lbrace \operatorname{Re} \left[ \lambda_{\ell}\right] \right\rbrace <0. \label{Eq50}
	\end{align}
	\begin{figure}[ht]
		\centering
		\includegraphics[width=0.48\textwidth]{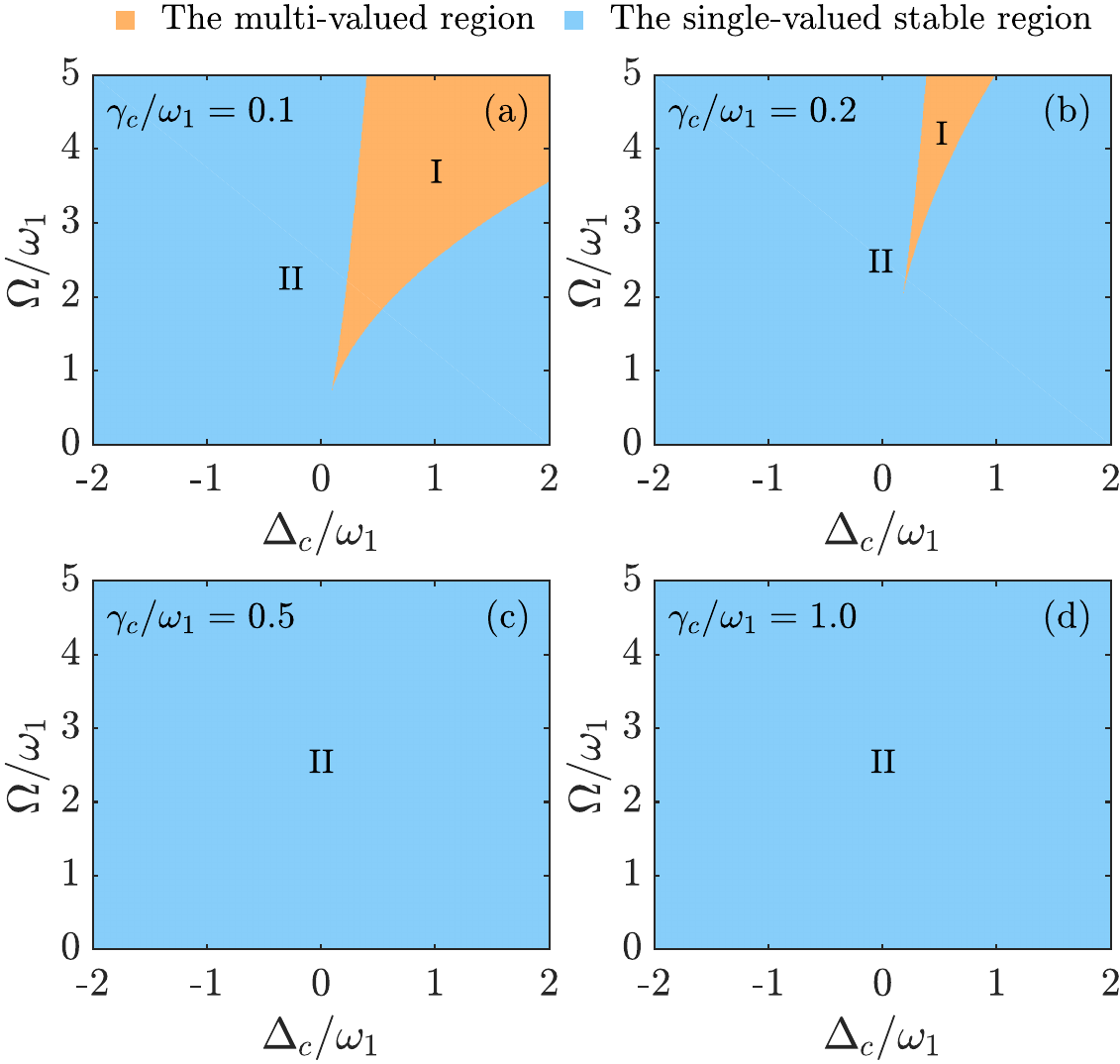}
		\caption{\label{fig:2}Stability phase diagram of the three-mode optomechanical system as a function of $\Delta_c/\omega_1$ and $\Omega/\omega_1$ when (a) $\gamma_c/\omega_1 = 0.1$, (b) $\gamma_c/\omega_1 = 0.2$, (c) $\gamma_c/\omega_1 = 0.5$, and (d) $\gamma_c/\omega_1 = 1.0$. The orange region (Region~I) denotes a multivalued semiclassical response, whereas the blue region (Region~II) denotes a unique linearly stable operating point within the RWA effective model. Other parameters used are $\chi = 1.18$, $g_1/\omega_1 =0.01$, $g_2/\omega_1 = 0.02$, $\gamma_1/\omega_1 =\gamma_2/\omega_1 = 2\times10^{-5}$, $\omega_\chi/\omega_1 = 5$, $n_0 = 1$, $\varphi_0 = 0$, and $\bar n_{\mathrm{th},1}=\bar n_{\mathrm{th},2}=0$.}
	\end{figure}
	
	To analyze the stability of the system, we solve Eq.~\eqref{Eq43} for each pair $(\Delta_c,\Omega)$ and retain its nonnegative real roots. The parameter space with multiple physical roots is assigned to Region~I and corresponds to a multivalued semiclassical response. Region~I is classified solely according to the number of physical roots; the individual branches are not further classified by their stability in these diagrams. When a unique physical root is obtained, we calculate the corresponding semiclassical amplitudes, construct $\mathbf A$, and evaluate its eigenvalues. All unique operating points found within the parameter ranges considered satisfy Eq.~\eqref{Eq50} and are assigned to Region~II. No single-valued unstable region is found within the RWA effective model. This result is consistent with the passive beam-splitter form of Eq.~\eqref{Eq24}, which permits coherent excitation exchange but provides no parametric amplification.
	
	Figure~\ref{fig:2} shows that a multivalued region occurs at positive cavity detuning and intermediate drive strengths for $\gamma_c/\omega_1=0.1$ and $0.2$. This behavior originates from the dependence of the effective cavity detuning $\bar{\Delta}_c'$ on the intracavity photon number, which produces nonlinear optomechanical feedback. Increasing $\gamma_c$ reduces the intracavity field and thereby weakens this feedback. Consequently, Region~I contracts as $\gamma_c$ increases and is absent for $\gamma_c/\omega_1=0.5$ and $1.0$ over the parameter range shown. The parameters used in the subsequent analysis of the approximate Hamiltonian and dark-mode population dynamics are selected from	Region~II, ensuring a unique semiclassical working point that is linearly stable within the considered model.
	
	\section{Fidelity Assessment of the Approximate Hamiltonian}\label{sec:5}
	
	To assess the accuracy of the approximate Hamiltonian in Eq.~\eqref{Eq24}, we compare its dynamics with that generated by the time-dependent Hamiltonian $\hat H_1(t)$ in Eq.~\eqref{Eq19}. The latter retains the nonlinear radiation-pressure terms, counter-rotating terms, and nonselected Floquet sidebands omitted in the approximate Hamiltonian. The fidelity therefore quantifies the combined discrepancy arising from these approximations at a fixed semiclassical operating point. The semiclassical amplitudes entering both Hamiltonians are determined self-consistently from the dissipative stationary equations in Sec.~\ref{sec:4A} and held fixed during the evolution. Dissipations are omitted for both fluctuation evolutions. Hereafter, we omit $\delta$ from the fluctuation operators for brevity.
	
	Concretely, we assume that the three modes are prepared in the coherent states
	\begin{align}
		|\Psi(0)\rangle	=|\mu\rangle_a \otimes|\nu_{1}\rangle_{b_1}	\otimes|\nu_{2}\rangle_{b_2},\label{Eq51}
	\end{align}
	where $\mu$ and $\nu_{j}$ ($j=1,2$) are the initial	coherent-state amplitudes of the cavity field and mechanical fluctuation modes, respectively.
	
	To obtain the approximate state governed by the approximate Hamiltonian $\hat{H}_{\mathrm{app}}$, we write the	number-conserving quadratic Hamiltonian as $\hat H_{\mathrm{app}} =\hat{\boldsymbol c}^{\dagger}\mathbf M \hat{\boldsymbol c}$, with $\hat{\boldsymbol c}=(\hat a,\hat b_1,\hat b_2)^{\mathrm T}$ and
	\begin{align}
		\mathbf M= 
		\begin{pmatrix}
			0 & g_{1,\mathrm{eff}} & g_{2,\mathrm{eff}}\\
			g_{1,\mathrm{eff}}^* & -d_1 & 0\\
			g_{2,\mathrm{eff}}^* & 0 & -d_2
		\end{pmatrix}.
		\label{Eq52}
	\end{align}
	The corresponding Heisenberg equations involve only the annihilation operators, so the initial product of coherent states remains a product of coherent states~\cite{HeBergou2008}. The coherent amplitudes $\boldsymbol{z}(t)= \bigl(\mu(t),\nu_{1}(t),\nu_{2}(t)\bigr)^{\mathrm T}$ evolve as $\boldsymbol z(t)=e^{-i\mathbf M t}\boldsymbol z(0)$, where $\boldsymbol z(0)=(\mu,\nu_{1},\nu_{2})^{\mathrm T}$. The approximate state is given by
	\begin{align}
		|\Psi_{\mathrm{app}}(t)\rangle=	|\mu(t)\rangle_a \otimes|\nu_{1}(t)\rangle_{b_1}\otimes|\nu_{2}(t)\rangle_{b_2}. \label{Eq53}
	\end{align}
	
	To compute the reference evolution, we expand the state $|\Psi_1(t)\rangle$ evolved under $\hat H_1(t)$ in the three-mode Fock basis as
	\begin{align}
		|\Psi_1(t)\rangle =\sum_{n_a,n_1,n_2=0}^{\infty} C_{n_a,n_1,n_2}(t)|n_a\rangle_a\otimes|n_1\rangle_{b_1} \otimes|n_2\rangle_{b_2}.	\label{Eq54}
	\end{align}
	Here, $n_a$, $n_1$, and $n_2$ are the occupation numbers of the cavity field and the two mechanical modes, respectively. The coefficients $C_{n_a,n_1,n_2}(t)$ are the corresponding probability amplitudes, obtained by numerically integrating the Schr\"odinger equation for $\hat H_1(t)$. For the numerical calculation, the reference state is evolved in a truncated Fock space of dimension $\mathcal N=N_aN_{1}N_{2}$, where $N_a$, $N_{1}$, and $N_{2}$ are the truncation dimensions of the Fock spaces for the cavity field $a$ and the two mechanical modes $b_1$ and $b_2$, respectively.

	\begin{table}[htbp]
		\centering
		\caption{Fock-space convergence of the fidelity at $\omega_1t=10$ with $N_{\max}=6$. Here, $N_a$, $N_{1}$, and $N_{2}$ are the retained numbers of the Fock states, and $F(\mu)$ denotes the fidelity for the coherent state amplitude $\mu$. Other parameters used are the same as those in Fig.~\ref{fig:3}.}
		\label{tab:fock_coh}
		\begin{tabular*}{\columnwidth}
			{@{\extracolsep{\fill}}cccccc@{}}
			\hline\hline
			$N_a$ & $N_{1}$ & $N_{2}$ & $\mathcal N$ & $F(0.1)$ & $F(1.0)$\\
			\hline
			6  & 4 & 4 & 96  & 0.9999413762 & 0.9986113884\\
			8  & 5 & 5 & 200 & 0.9999413762 & 0.9986053706\\
			10 & 6 & 6 & 360 & 0.9999413762 & 0.9986041582\\
			14 & 8 & 8 & 896 & 0.9999413762 & 0.9986040918\\
			\hline\hline
		\end{tabular*}
	\end{table}
	
	\begin{table}[htbp]
		\centering
		\caption{Convergence of the fidelity at $\omega_1t=10$ with respect to the Jacobi--Anger sideband order cutoff number $N_{\max}$. The Fock-space dimensions are $(N_a,N_{1},N_{2})=(8,5,5)$. The notation and other parameters are the same as in Table~\ref{tab:fock_coh}.}
		\label{tab:jacobi_coh}
		\begin{tabular*}{\columnwidth}
			{@{\extracolsep{\fill}}cccc@{}}
			\hline\hline
			$N_{\max}$ & $F(0.1)$ & $F(1.0)$ & $|J_{N_{\max}+1}(1.18)|$\\
			\hline
			2  & 0.9999426726 & 0.9986154021 & $3.135\times10^{-2}$\\
			4  & 0.9999413628 & 0.9986052840 & $5.621\times10^{-4}$\\
			6  & 0.9999413762 & 0.9986053706 & $4.727\times10^{-6}$\\
			8  & 0.9999413762 & 0.9986053702 & $2.305\times10^{-8}$\\
			10 & 0.9999413762 & 0.9986053702 & $7.338\times10^{-11}$\\
			12 & 0.9999413762 & 0.9986053702 & $1.644\times10^{-13}$\\
			\hline\hline
		\end{tabular*}
	\end{table}
	To compare the two states in the same frame, we apply $\hat U^\dagger(t)$ from Eq.~\eqref{Eq23} to obtain $|\Psi_{\mathrm{ref}}(t)\rangle =\hat U^\dagger(t)|\Psi_1(t)\rangle$ in the frame corresponding to the approximate Hamiltonian. The fidelity is defined as
	\begin{align}
		F(t)= \left|\langle\Psi_{\mathrm{ref}}(t)|\Psi_{\mathrm{app}}(t)\rangle \right|^2. \label{Eq55}
	\end{align}
	The reference evolution starts from the normalized projection of Eq.~\eqref{Eq51}. The analytic approximate state is projected onto the same space and normalized before evaluating the overlap, ensuring $F(0)=1$. The Jacobi--Anger expansion is truncated to $-N_{\max}\leq n\leq N_{\max}$. The semiclassical background is recalculated at each parameter point. Both $C$ and $\alpha_{\mathrm{ss}}$ are updated when varying $\chi$.
	
	To assess the numerical convergence	of the fidelity, we evaluate the fidelity by varying the Fock-space truncation dimensions and the Jacobi--Anger cutoff number. Tables~\ref{tab:fock_coh} and \ref{tab:jacobi_coh} summarize the results at $\omega_1t=10$ for $\nu_1=\nu_2=0.5\mu$. For $\mu=0.1$, the fidelities obtained with all four Fock-space truncations agree to agree to the ten decimal places reported in Table~\ref{tab:fock_coh}. For $\mu=1.0$, the results obtained with $(N_a,N_1,N_2)=(8,5,5)$ and $(14,8,8)$ differ by approximately $1.3\times10^{-6}$, whereas the difference between the two largest	truncations is approximately $6.6\times10^{-8}$. At fixed dimensions $(8,5,5)$, increasing $N_{\max}$ from $6$ to $12$ changes the fidelity by less than $5\times10^{-10}$. For $\mu=1.0$, the calculated fidelity is therefore more sensitive to the Fock-space truncation than to the Jacobi--Anger cutoff over the ranges tested, but the observed changes are much smaller than the infidelity discussed below. The calculations below use	$(N_a,N_1,N_2)=(8,5,5)$ and $N_{\max}=6$. These tests apply to the specified time and initial amplitudes.
	
\begin{figure}[t]
		\centering
		\includegraphics[width=\linewidth]{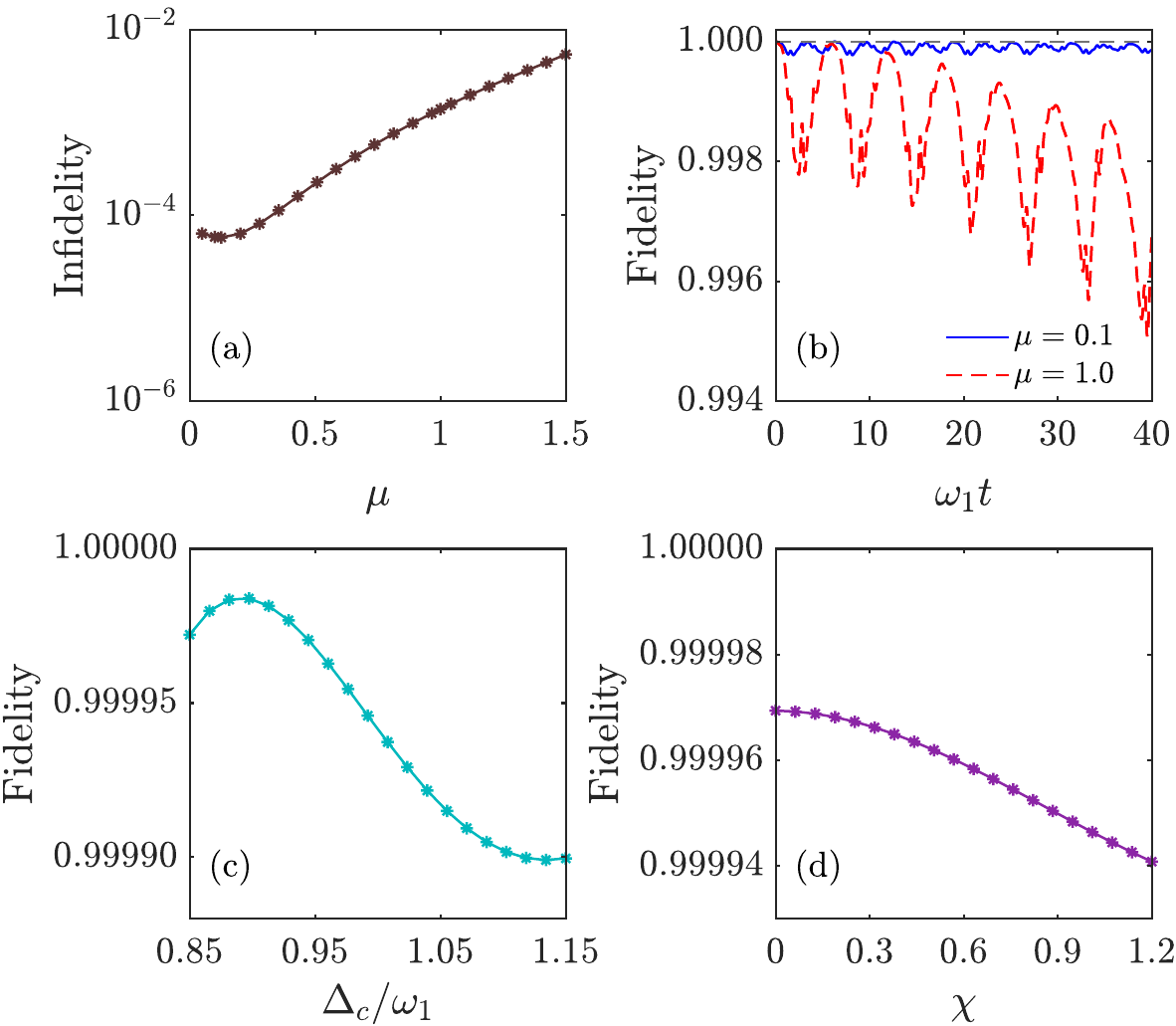}
		\caption{\label{fig:3}Fidelity between the reference and approximate states when the system is in an initial three-mode coherent state. (a) Infidelity $1-F$ at $\omega_1t=10$ as a function of $\mu$.	(b) Fidelity as a function of $\omega_1t$ for $\mu=0.1$ (blue solid line) and $1.0$ (red dashed line). The gray dashed line marks $F=1$.	(c) Fidelity at $\omega_1t=10$ as a function of the bare cavity detuning $\Delta_c/\omega_1$. (d) Fidelity at $\omega_1t=10$ as a function of the dimensionless modulation depth $\chi$. Here, $\Delta_c/\omega_1=1.0004$ in (a), (b), and (d), $\chi=1.18$ in (a)--(c), and $\mu=0.1$ in (c) and (d). In all panels, $\nu_{1}=\nu_{2} =0.5\mu$, $\gamma_c/\omega_1=0.2$, and $\Omega/\omega_1=1$. Other parameters used are the same as in Fig.~\ref{fig:2}. The numerical cutoffs are $(N_a,N_1,N_2)=(8,5,5)$ and $N_{\max}=6$.}
	\end{figure}

	We also examine the accuracy of the approximate Hamiltonian	for different initial coherent-state amplitudes and evolution times. Figure~\ref{fig:3}(a) shows that increasing $\mu$ generally increases the infidelity at $\omega_1t=10$, apart from a weak nonmonotonic variation at small amplitudes. The fidelities for $\mu=0.1$ and $1.0$ are approximately $0.999941$ and $0.998605$, respectively. The amplitude dependence is also evident in the time evolution shown in Fig.~\ref{fig:3}(b): the fidelity for $\mu=0.1$ remains close to unity, whereas that for $\mu=1.0$ exhibits stronger oscillations and larger deviations from unity. Thus, the approximate Hamiltonian describes the smaller-amplitude evolution more accurately over the interval examined.
	
	To examine the dependence of the fidelity on the system parameters, we compare the fidelities at $\omega_1t=10$ for $\mu=0.1$, varying the bare cavity detuning and modulation depth separately. Figure~\ref{fig:3}(c) shows a nonmonotonic dependence on the bare cavity detuning over $0.85\leq\Delta_c/\omega_1\leq1.15$, with a maximum near $\Delta_c/\omega_1=0.90$. The minimum fidelity over this range is approximately $0.999899$. With the bare cavity detuning fixed at its reference value, Fig.~\ref{fig:3}(d) shows that the fidelity decreases from approximately $0.999969$ to $0.999941$ as $\chi$ increases from 0 to 1.2. Varying $\chi$ changes both the Floquet sideband coefficients and the self-consistent semiclassical steady-state amplitudes, so this scan also probes changes in the effective couplings and residual detunings.	The high fidelity is therefore maintained across both scanned parameter ranges at the chosen time and initial amplitude. These results support the validity of the approximate Hamiltonian for the parameters used.
	
	\section{Dark-mode population routing and controlled release}\label{sec:6}
	
	To exhibit the physical principle of the dark-mode manipulation, we design a scheme to preserve mechanical excitations and release them on demand. We therefore examine how mechanical frequency mismatch controls excitation transfer from the dark mode to the cavity mode through the bright mode. We begin with the beam-splitter Hamiltonian in Eq.~\eqref{Eq26}, obtained under the single-photon resonance condition in Eq.~\eqref{Eq25}. We introduce the collective mechanical modes	\cite{Shkarin2014,Dong2012,WangClerk2012NJP}
	\begin{subequations}\label{Eq56}
		\begin{align}
			\hat{B}_{+}	&=\frac{g_{1,\mathrm{eff}}\hat{b}_1 + g_{2,\mathrm{eff}}\hat{b}_2}{G},	\label{Eq56a}\\
			\hat{B}_{-} &=\frac{g_{2,\mathrm{eff}}^*\hat{b}_1-g_{1,\mathrm{eff}}^*\hat{b}_2}{G}, \label{Eq56b}
		\end{align}
	\end{subequations}
	where we introduce the parameter
	\begin{align}
		G=\sqrt{|g_{1,\mathrm{eff}}|^2+	|g_{2,\mathrm{eff}}|^2}.\label{Eq57}
	\end{align}
	These operators satisfy $[\hat{B}_{\pm},\hat{B}_{\pm}^{\dagger}]=1$ and	$[\hat{B}_{+},\hat{B}_{-}^{\dagger}]=0$. Based on Eqs.~\eqref{Eq56}, we can obtain the inverse transformation as
	\begin{subequations}\label{Eq58}
		\begin{align}
			\hat{b}_1 &= \frac{g_{1,\mathrm{eff}}^*\hat{B}_{+}+g_{2,\mathrm{eff}}\hat{B}_{-}}{G}, \label{Eq58a}\\
			\hat{b}_2 &= \frac{g_{2,\mathrm{eff}}^*\hat{B}_{+}-g_{1,\mathrm{eff}}\hat{B}_{-}}{G}. \label{Eq58b}
		\end{align}
	\end{subequations}
	Substituting Eqs.~\eqref{Eq58} into Eq.~\eqref{Eq26} gives
	\begin{align}
		\hat{H}_{\mathrm{app}}'=G(\hat{a}^{\dagger}\hat{B}_{+} +\hat{a}\hat{B}_{+}^{\dagger}). \label{Eq59}
	\end{align}
	Equation~\eqref{Eq59} shows that $\hat B_+$ couples to the cavity mode $a$ with strength $G$, whereas $\hat B_-$ is decoupled from the cavity field. We identify $\hat B_+$ and $\hat B_-$ as the bright and dark mechanical modes, respectively. For the equal mechanical damping rates used here, the dark mode is protected from cavity-induced decay but remains subject to intrinsic mechanical damping.
	
	\begin{figure}[ht]
		\centering
		\includegraphics[width=\linewidth]{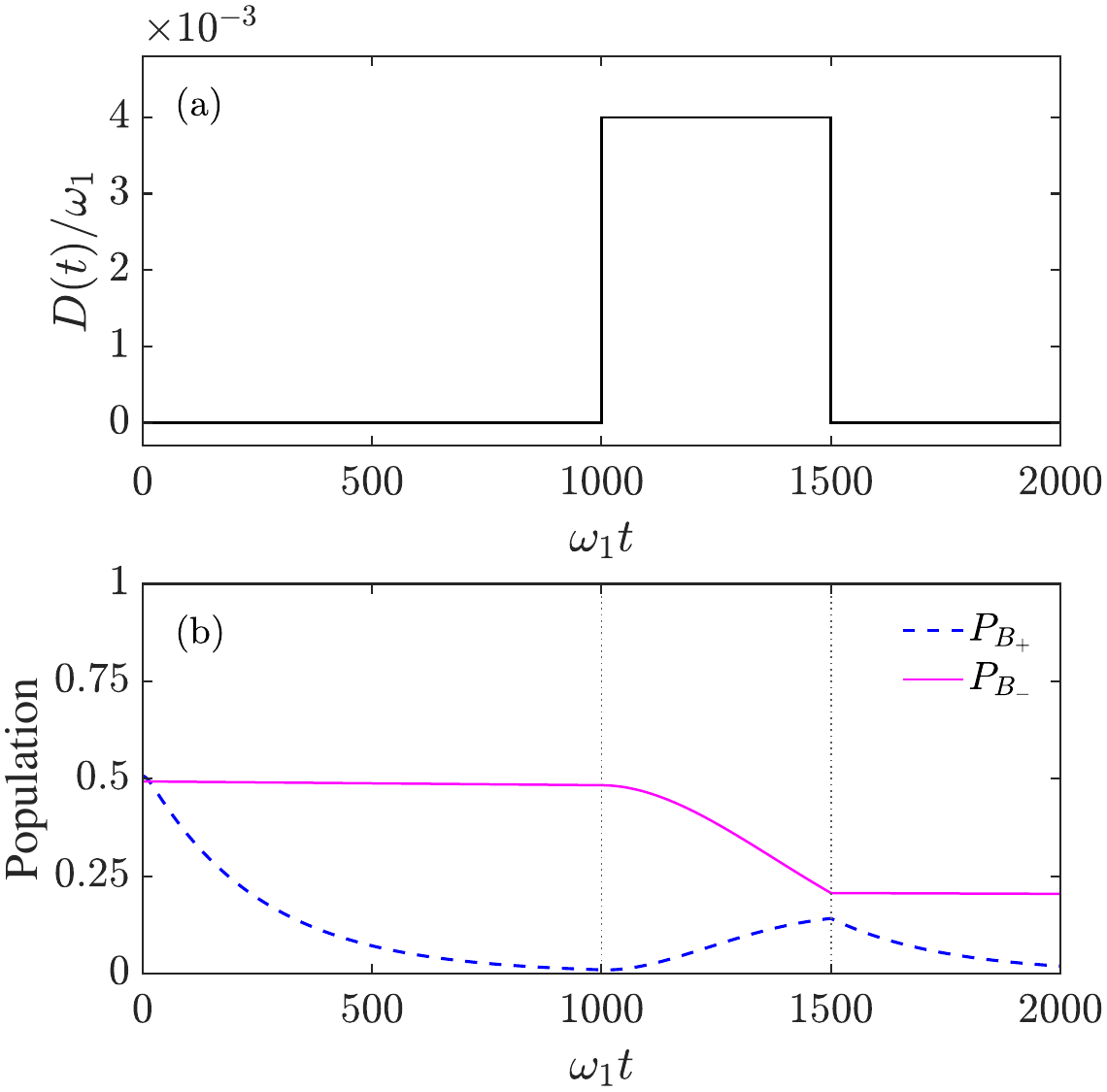}
		\caption{\label{fig:4}Detuning-controlled population dynamics of the bright and dark modes.	(a) Mechanical detuning $D(t)/\omega_1$	as a function of $\omega_1t$. The detuning is $D_0/\omega_1=0.004$ during $1000\leq\omega_1t<1500$ and zero otherwise. (b) Populations $P_{B_+}$ (blue dashed line) and $P_{B_-}$ (magenta solid line) of the reference bright and dark modes, respectively. The vertical dotted lines mark the beginning and end of the detuning pulse. The parameters used are $\Delta_c/\omega_1=1.0004$, $\gamma_c/\omega_1=0.2$, and $\Omega/\omega_1=1$. Other parameters used are the same as in Fig.~\ref{fig:2}.}
	\end{figure}
	We consider a single excitation initially localized in the first mechanical mode with the state given by
	\begin{align}
		|\psi(0)\rangle	=|0\rangle_a|1\rangle_{b_1}|0\rangle_{b_2}.	\label{Eq60}
	\end{align}
	In the bright-dark-mode representation, the mechanical part is given by
	\begin{align}
		|1\rangle_{b_1}|0\rangle_{b_2} &= \frac{g_{1,\mathrm{eff}}}{G}|1\rangle_{B_+}|0\rangle_{B_-}+\frac{g_{2,\mathrm{eff}}^*}{G}|0\rangle_{B_+}|1\rangle_{B_-}.\label{Eq61}
	\end{align}
	The initial populations of the bright and dark modes are therefore given by $P_{B_+}(0)=|g_{1,\mathrm{eff}}|^2/G^2$ and $P_{B_-}(0)=|g_{2,\mathrm{eff}}|^2/G^2$, respectively. The bright-mode excitation can dissipate through the cavity, whereas the dark-mode excitation is protected from cavity-induced decay under frequency matching.
	
	\begin{figure}[ht]
		\centering
		\includegraphics[width=\linewidth]{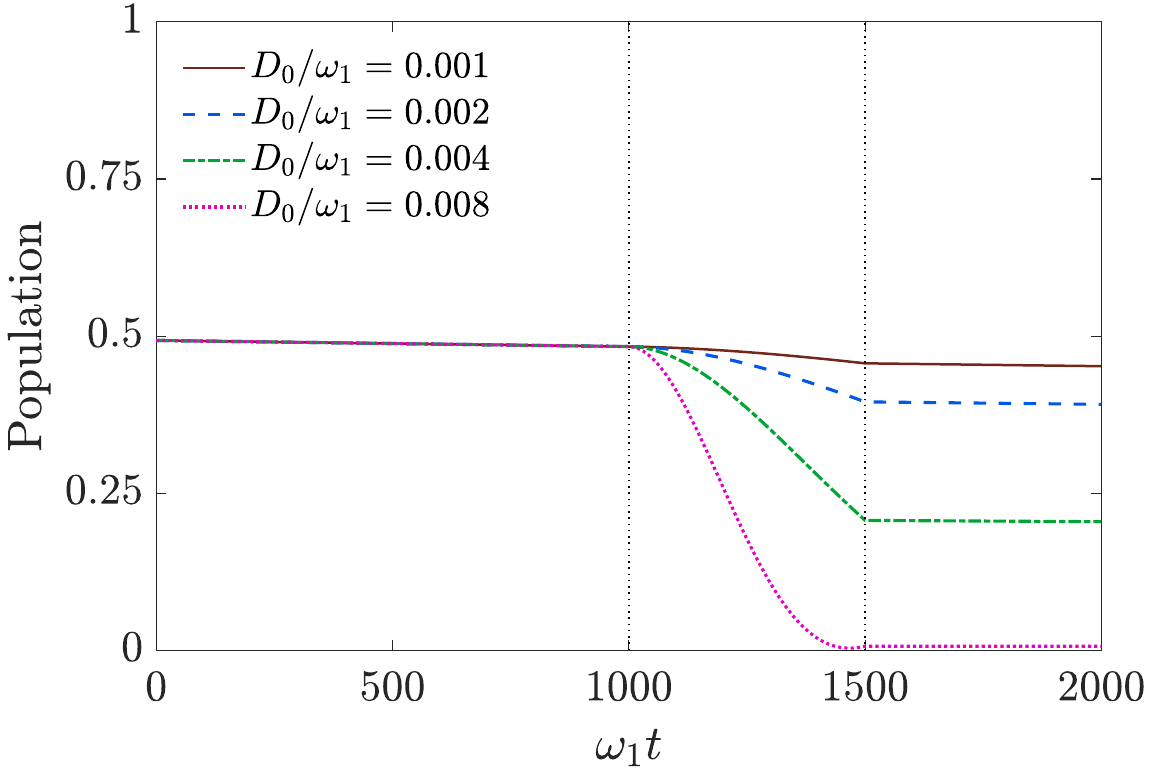}
		\caption{\label{fig:5}Dynamics of the dark-mode population when the effective mechanical detuning amplitude takes different values: $D_0/\omega_1=0.001$, $0.002$, $0.004$, and $0.008$. Each pulse is applied during
			$1000\leq\omega_1t<1500$, with $D(t)/\omega_1=0$ outside this interval. Other parameters used are the same as those in Fig.~\ref{fig:4}.}
	\end{figure}
	To control the dark-mode population dynamics, we introduce a time-dependent frequency mismatch $D(t)=\delta_2(t)-\omega_1$ between the effective frequencies of the two mechanical modes. Equation~\eqref{Eq24} can then be written as
	\begin{align}
		\hat{H}_{\mathrm{app}}(t) &=g_{1,\mathrm{eff}}\hat{a}^{\dagger}\hat{b}_1 + g_{2,\mathrm{eff}}\hat{a}^{\dagger}\hat{b}_2 +\mathrm{H.c.} \notag\\
		&\quad-d_1(\hat{b}_1^{\dagger}\hat{b}_1+\hat{b}_2^{\dagger}\hat{b}_2)+D(t)\hat{b}_2^{\dagger}\hat{b}_2.	\label{Eq62}
	\end{align}
	The effective couplings are evaluated at the reference steady-state operating point and held fixed throughout the detuning pulse. Substituting the inverse transformation \eqref{Eq58} into Eq.~\eqref{Eq62}, we obtain the following expression for the Hamiltonian in the bright-dark-mode representation
	\begin{align}
		\hat{H}_{\mathrm{app}}(t) &=G(\hat{a}^{\dagger}\hat{B}_{+} + \hat{a}\hat{B}_{+}^{\dagger}) - d_1(\hat{B}_{+}^{\dagger}\hat{B}_{+} +	\hat{B}_{-}^{\dagger}\hat{B}_{-}) \notag\\
		&\quad+\frac{D(t)}{G^2} \Bigl[|g_{2,\mathrm{eff}}|^2	\hat{B}_{+}^{\dagger}\hat{B}_{+} + |g_{1,\mathrm{eff}}|^2 \hat{B}_{-}^{\dagger}\hat{B}_{-} \notag\\
		&\quad-g_{1,\mathrm{eff}}g_{2,\mathrm{eff}}\hat{B}_{+}^{\dagger}\hat{B}_{-} - g_{1,\mathrm{eff}}^*g_{2,\mathrm{eff}}^*\hat{B}_{-}^{\dagger}\hat{B}_{+}\Bigr]. \label{Eq63}
	\end{align}
	The off-diagonal terms describe the excitation exchange between the reference bright and dark modes, with coupling coefficient $\lambda(t)=-D(t)g_{1,\mathrm{eff}}g_{2,\mathrm{eff}}/G^2$.
	A finite frequency mismatch therefore couples the dark mode to the cavity-coupled bright mode, opening a pathway for dissipation through the cavity field. The diagonal terms proportional to $D(t)$ also shift the collective-mode frequencies, so the release dynamics depends on both these shifts and the induced coupling.	When $D(t)/\omega_1=0$, the bright-dark-mode coupling vanishes; the common residual 
	detuning $d_1$ does not mix the two modes. Restoring the frequency matching thus suppresses further transfer from the dark mode, although the excitations in the bright mode or the cavity field can continue to dissipate.
	
	In the numerical simulations, we calculate the populations $P_{B_\pm}(t) 	=\langle\hat B_\pm^\dagger\hat B_\pm\rangle$ using the fixed collective-mode operators defined in Eq.~\eqref{Eq56}. These operators remain unchanged during the pulse, so $P_{B_\pm}(t)$ describe the populations of the same reference modes even when the detuning couples them.
	
	We examine whether a detuning pulse can induce	dark-mode population release and whether restoring frequency matching suppresses further transfer to the bright mode. Figure~\ref{fig:4}(a) shows the mechanical detuning, with $D(t)=D_0$ during $1000\leq\omega_1t<1500$ and $D(t)=0$ otherwise, where $D_0/\omega_1=0.004$. The corresponding population dynamics is shown in Fig.~\ref{fig:4}(b). The initial state in Eq.~\eqref{Eq60} contains both bright and dark components, with $P_{B_+}(0)\simeq0.507$ and $P_{B_-}(0)\simeq0.493$ for the parameters considered. Before the pulse, the bright component couples to the lossy cavity and decays, whereas the dark component remains decoupled from the subsystem formed by the bright mode and the cavity mode. The dark-mode population decreases slowly through intrinsic mechanical damping. During the pulse, the finite detuning activates the bright--dark-mode coupling in Eq.~\eqref{Eq63}. The decrease in $P_{B_-}$ and temporary increase in $P_{B_+}$ indicate excitation transfer from the reference dark mode to the cavity-coupled bright mode, allowing the transferred excitation to dissipate through the cavity loss. Removing the detuning at $\omega_1t=1500$ restores frequency matching and switches off this coupling. The remaining dark-mode population resumes its slow decay under mechanical damping, while the excitation already in the bright mode or the cavity field can continue to dissipate. Restoring frequency matching thus suppresses further transfer from the dark mode and protects the surviving dark-mode population.
	
	To examine how the population release depends on the detuning amplitude, we calculate $P_{B_-}(t)$ for $D_0/\omega_1=0.001$, $0.002$, $0.004$, and $0.008$, keeping the effective couplings, damping rates, initial state, and pulse duration unchanged.	As shown in Fig.~\ref{fig:5}, the four curves coincide before the pulse and separate once the detuning is applied. For the amplitudes considered, the population remaining at $\omega_1t=1500$ decreases monotonically with increasing $D_0$. The smallest amplitude, $D_0/\omega_1=0.001$, produces only weak depletion, whereas $D_0/\omega_1=0.008$ nearly depletes the reference dark-mode population during the same pulse interval. After the frequency matching is restored, the remaining populations resume their slow decay. The pulse amplitude therefore controls the population remaining in the reference dark mode for the fixed duration and parameter range examined.
	
	Figures~\ref{fig:4} and \ref{fig:5} demonstrate the detuning-controlled preservation and release of dark-mode population within the approximate three-mode model. A finite effective mechanical frequency mismatch enables excitation transfer to the cavity-coupled bright mode,	while restoring frequency matching suppresses further transfer and protects the remaining dark-mode population. In addition, for the initial state used in Figs.~\ref{fig:4} and \ref{fig:5}, we numerically compare the reference and approximate evolutions with dissipation and the detuning pulse omitted. The fidelity between the two states remains at or above $0.934$ throughout the interval	 $0\leq\omega_1t\leq2000$.
	
	\section{Conclusion}\label{sec:7}
	
	In conclusion, we have proposed a scheme to realize the Floquet dark-mode manipulation in a three-mode optomechanical system. By selecting a sideband generated by mechanical frequency modulation, we derived an approximate beam-splitter Hamiltonian supporting collective bright and dark modes under effective mechanical frequency matching. This	description was evaluated at a self-consistent semiclassical operating point in a single-valued regime, where the fluctuation dynamics is linearly stable for the model under the rotating-wave approximation. Fidelity comparisons with the time-dependent dynamics, together with numerical convergence tests, support its validity under the conditions examined. Within this framework, a finite effective mechanical detuning breaks the dark mode such that it can be coupled to the cavity-coupled bright mode, enabling a population release through the cavity-loss channel. Restoring the frequency matching suppresses	further transfer and protects the remaining dark-mode population. The amplitude scan further shows that, for a fixed pulse duration, the release can be varied from a weak depletion to nearly complete depletion of the reference dark-mode population over the range considered. Our method provides a reliable means to manipulate the dark-mode effect in coupled bosonic mode systems.
	
	\begin{acknowledgments}
		J.-Q.L. thanks Prof. Jiteng Sheng for valuable discussions. J.-Q.L. was supported in part by the National Natural Science Foundation of China (Grants No. 12575015, No. 12247105, and No. 12421005), National Key Research and Development Program of China (Grant No. 2024YFE0102400), and Hunan Provincial Major Sci-Tech Program (Grant No. 2023ZJ1010). J.-F. H. is partially supported by the National Natural Science Foundation of China (Grant No. 12475016) and Young Talents of the Furong Program of Hunan Province (Grant No. 2025RC3141).
	\end{acknowledgments}

\end{document}